# Imputing unknown competitor marketing activity with a Hidden Markov Chain

Dominique Haughton, Bentley University, Université Paris I (Panthéon-Sorbonne) and Université Toulouse I[1]
Guangying Hua, Boston Consulting Group
Danny Jin, Epsilon
John Lin, Epsilon
Qizhi Wei, Epsilon
Changan Zhang, Bentley University[2]

*Abstract:*

We demonstrate on a case study with two competing products at a bank how one can use a Hidden Markov Chain (HMC) to estimate missing information on a competitor's marketing activity. The idea is that given time series with sales volumes for products A and B and marketing expenditures for product A, as well as suitable predictors of sales for products A and B, we can infer at each point in time whether it is likely or not that marketing activities took place for product B. The method is successful in identifying the presence or absence of marketing activity for product B about 84% of the time. We allude to the issue of whether, if one can infer marketing activity about product B from knowledge of marketing activity for product A and of sales volumes of both products, the reverse might be possible and one might be able to impute marketing activity for product A from knowledge of that of product B. This leads to a concept of symmetric imputation of competing marketing activity. The exposition in this paper aims to be accessible and relevant to practitioners.

[1] Supported by NSF DMS-1106388
[2] Supported by NSF DMS-1106388

# Introduction

In many cases, it is of interest to estimate the impact of a promotion activity on sales volume of a product. Many efforts to optimize the media mix for a product face this issue. Typically a time series is available with sales volume and promotion activity data at the weekly or monthly level, and at various levels of geographical disaggregation. In a number of cases, data on sales volume for a competitor are available, but data on that competitor's promotion activity is not available. We note that this state of affairs may depend on the geographical location of the market of interest: for example, in the United Kingdom, extensive market research provides share of voice and other marketing and sales metrics across most players in any market. In the United States such data may exist but be difficult and very costly to obtain.

The objective of this paper is to demonstrate a technique for imputing a competitor's promotion activity, with two goals in mind:

1. To obtain a more accurate estimate of the impact of a promotion activity on sales volume (by incorporating in the model an imputed value for competing marketing activity)
2. To track competing marketing activity and act in consequence

Our main contributions are:

1. To demonstrate the feasibility and desirability of the method described here
2. To provide a roadmap, inclusive of suitable code, to practitioners who might wish to implement it

The approach we adopt in this paper, referred to as the "Hidden Markov Model (HMM)" approach, was introduced by Moon et al. (2007) and Ledolter (2007) in the context of two competing pharmaceutical products; for instance one might imagine two competing allergy drugs. The idea is to model the unknown competing marketing activity as a latent Markov chain which takes values 0 (no competitor activity) or 1 (competitor activity) at each time period *t*. In this paper, as in Moon et al., this latent Markov chain is assumed to be of order 1. This implies that given past competitor activity, the probability of competitor activity taking place at time *t* depends on whether competitor activity took place at time *t-1* but not on whether it took place at earlier time periods. In a number of applications, this assumption will be at least approximately correct.

Consider the case of two competing brands A and B and assume that one of the brands, say A, is the focal brand. Assume that:

1. The brand manager of the focal brand A has data on promotion activity for brand A
2. The manager also has sales volume information for both brands A and B

In order to estimate the impact of promotion activity for Brand A, often with an objective of identifying a better marketing mix, the manager of brand A wishes to build a time series model for its sales volume in terms of its promotion activity over several channels. This method is commonly used in many marketing mix applications.

The problem is that ignoring a competitor's promotion activity may lead to biased estimates of the effect of one's own promotion efforts, in turn leading to suboptimal marketing mix decisions.

Moon et al. (2007), and Ledolter (2007) show with the help of synthetic (simulated) data that the model successfully imputes unknown competitor activity and reduces biases in the estimation of the impact of the promotion activity for the focal product on sales. The authors then apply the model to actual pharmaceutical data. Moon at al. (2007) employ a linear model, while Ledolter (2007) implements a Poisson regression model; both papers reach similar conclusions. We note that the data in both Moon et al. (2007) and Ledolter (2007) are available at the physician level. In order to estimate the latent states of the Markov chain, MCMC (Monte Carlo Markov Chain) techniques are used in both papers. Ledolter employs the software Winbugs to generate posterior distributions of parameters of interests, as will be done here.

## The data

To demonstrate the feasibility of the HMM approach, we will use the following dataset, concerning a bank in the United States. The name of the bank is omitted for confidentiality reasons, but we have data at hand that were used for a media mix modeling project. The dataset contains weekly data for 156 periods from 7/1/2006 to 6/30/2009 for the variables listed in Table 1, including:

1. new checking accounts and promotion data
2. new MMDAs - Money Market Deposit Accounts
3. promotion data (marketing expenditures)

*Table 1. List of variables and summary statistics (sample size = 156)*

| Variable | Minimum | Maximum | Mean | Std. Dev. |
|---|---|---|---|---|
| # New checking accounts | 1,320.00 | 14,347.00 | 3,429.21 | 2,209.70 |
| # New MMDA accounts | 324.00 | 1,362.00 | 686.42 | 201.97 |
| Total marketing expenditures for checking accounts (Newspaper + Radio + Internet + outdoor) in US $ | 0.00 | 578,091.21 | 83,487.32 | 143,055.45 |
| Total marketing expenditures for MMDA (Newspaper + Radio) in US $ | 0.00 | 315,912.43 | 69,826.03 | 106,514.54 |
| Dow Jones Industrial Average (DJIA) | 6,749.88 | 14,019.70 | 11,436.81 | 1,973.46 |
| Seasonality | 1,517.00 | 3,281.00 | 2,485.21 | 353.09 |
| Gift promotion (yes 1, no 0) | 0.00 | 1.00 | 0.05 | 0.22 |

Table 1 displays the list of variables used in this paper, along with summary statistics. The gift promotion variable takes the value of 1 if a promotional gift was offered that week (details on the promotional gift are omitted for confidentiality reasons) and 0 if not. The seasonality variable is specific to the model for new checking accounts and is computed by averaging past (pre 7/1/2006) historical valid numbers of new checking accounts for each of the 52 weeks in the year, excluding erroneous values.

## The approach to imputing competitor activity

To evaluate the HMM approach, we pretend that the MMDA promotion activity is unknown, estimate it with a Hidden Markov Model and evaluate the performance of the imputed promotion activity by comparing it with actual promotion activity. The model is defined by the following equations, where the subscript *t* for time periods has been omitted:

$$Y_{New\,Checking} = \alpha_{0c} + \alpha_{1c} Seasonality + \alpha_{2c} Gift + e_{New\,Checking}$$

$$Y_{New\,MMDA} = \alpha_{0m} + \alpha_{1m} DJIA + e_{New\,MMDA}$$

$$e_{New\,Checking} = \beta_{0c} + \beta_{1c} Z_{Checking\,Marketing\,Expenditure} + \beta_{2c} X_{MMDA\,Marketing\,Expenditure} + \varepsilon_{New\,Checking}$$

$$e_{New\,MMDA} = \beta_{0m} + \beta_{1m} Z_{Checking\,Marketing\,Expenditure} + \beta_{2m} X_{MMDA\,Marketing\,Expenditure} + \varepsilon_{New\,MMDA},$$

where $Z_{Checking\,Marketing\,Expenditure}$ is the known total amount of marketing expenditures for checking accounts and $X_{MMDA\,Marketing\,Expenditure}$ is the assumed unknown latent state of marketing activity for MMDAs, 1 if any activity took place (expenditures occurred at a given time period), 0 if not. Recall that while we do know the level of MMDA marketing expenditures, we pretend that we don't and represent those levels as an unknown 0/1 state to be estimated by the model. The terms $Y_{New\,Checking}$ and $Y_{New\,MMDA}$ represent the logarithms of the numbers of new checking accounts and MMDAs respectively.

The idea behind the equations is that first the effect of variables such as *Seasonality*, *Gift* and *DJIA* are removed from the time series of volumes of new checking accounts and MMDAs in the two first equations, and that the residuals resulting from this step are then modeled as in the last two equations. Figure 1 displays the model graphically as it applies to three consecutive periods *t, t+1,* and *t+2*. In Figure 1, $e_{t,1}$ refers to the residuals for new checking accounts (in the third equation) at time period *t*, $e_{t,2}$ refers to the residuals for new checking MMDAs (in the fourth equation) at time period *t* and t.p.m. is short for the Transition Probability Matrix which governs the Markov chain $X_t$ of MMDA marketing activity.

The output from linear regressions in the two first equations are given in Tables 2a (new checking accounts) and Table 2b (new MMDAs). The R-squares are respectively .631 and .291 and all variables are significant with p-values equal to zero to three decimal places. Both unstandardized and standardized regression coefficients are given since the scales of the independent variables differ widely.

*Table 2a. Regression results for new checking accounts (R-square = .631)*

| Dependent Variable: New checking accounts | Unstandardized coefficient | Standardized coefficient | p-value |
|---|---|---|---|
| Constant | 6.632 | | .000 |
| Seasonality | .001 | .427 | .000 |
| Gift | 1.171 | .587 | .000 |

*Table 2b. Regression results for new MMDAs (R-square = .291)*

| Dependent Variable: New checking accounts | Unstandardized coefficient | Standardized coefficient | p-value |
|---|---|---|---|
| Constant | 5.563 | | .000 |
| Seasonality | .00008 | .540 | .000 |

*Figure 1: Graphical representation of the model to impute missing competitor activity*

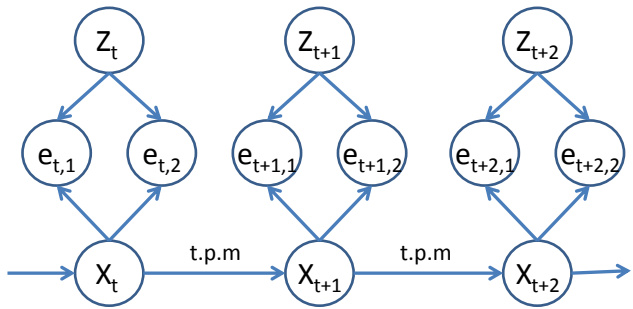

## Results

The first two equations in our model are estimated via a standard linear regression analysis and the residuals are extracted. The model in the two last equations is estimated via a Bayesian methodology, using the software Winbugs. Annotated Winbugs code is provided in the Appendix. Prior probability distributions are specified for all parameters to be estimated, including the transition probabilities for the Markov chain of MMDA marketing expenditures, the variances of the residual terms in the last two equations, as well as the covariance between these two residuals (in equations 3 and 4), and the $\beta$ coefficients.

Combining the prior distributions for the parameters with the likelihood of the data, following the Bayesian paradigm, yields the posterior distribution of the parameters, which is the distribution of the parameters conditioned on the data, and can be understood as an update of the prior distribution after observing the data. Typically the mean of the posterior distribution is used as an estimator for the parameter.

The density of posterior distributions cannot be computed in closed form in the vast majority of cases, so that obtaining posterior means and/or shapes of posterior distributions typically relies on simulating the posterior distribution, mostly via Monte Carlo Markov Chain (MCMC) methods. This is the approach taken by the software Winbugs.

Table 3 and Figure 2 illustrate the end result of the Bayesian procedure for the transition probabilities of the Markov chain on marketing activity for the MMDAs. For example we can see that the estimated probability of a marketing activity taking place at time period *t+1* given that none took place at time period *t* is about 12%, with a standard deviation of about 6 percentage points.

***Table 3: Posterior summary statistics for the components of the transition probability matrix***

| node   | mean    | sd      | 2.50%   | median  | 97.50% |
|--------|---------|---------|---------|---------|--------|
| P[1,1] | 0.8765  | 0.0633  | 0.7195  | 0.8888  | 0.9649 |
| P[1,2] | 0.1235  | 0.0633  | 0.03511 | 0.1112  | 0.2805 |
| P[2,1] | 0.04997 | 0.02772 | 0.01204 | 0.04469 | 0.119  |
| P[2,2] | 0.95    | 0.02772 | 0.881   | 0.9553  | 0.988  |

Figure 2 illustrates the fact that while posterior means give a Bayesian estimate for parameters of interest, the MCMC simulation provides a view of the whole posterior distribution, represented by kernel densities, which can be understood as smoothed versions of histograms.

One might wonder whether the correlation between new checking and MMDA accounts has been sufficiently addressed by the model, notably by the dependence of both time series upon the Markov chain of latent MMDA marketing activity. Table 4 and Figure 3 display posterior means and standard deviations of residual correlations, as well as residual variances. In Table 4, sigma[1,1] and sigma[2,2] represent residual standard deviations, and sigma[1,2] (or sigma[2,1]) the correlation between residuals for new checking accounts and new MMDA accounts. An interval defined by the 2.50[th] and 97.50[th] percentiles of the posterior distribution for that correlation includes 0, a Bayesian way of concluding that this correlation is not significantly different from zero.

*Figure 2: Posterior kernel densities for the components of the transition probability matrix*

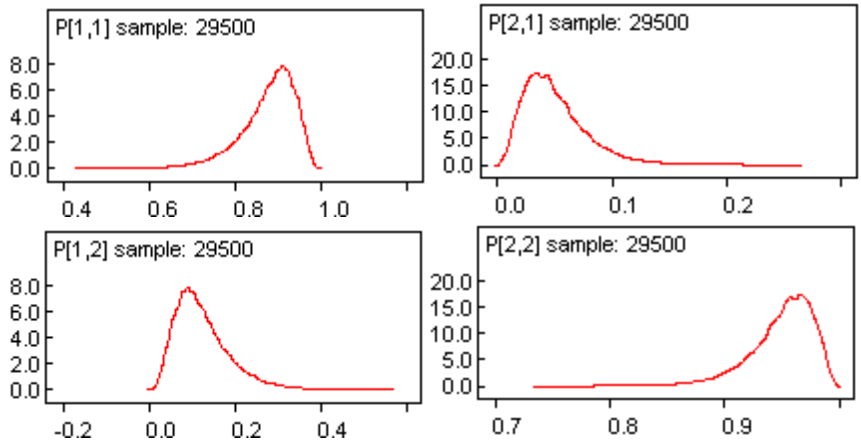

*Table 4: Posterior statistics for the variances and covariance of the random errors*

| node | mean | sd | 2.50% | median | 97.50% |
| --- | --- | --- | --- | --- | --- |
| sigma[1,1] | 0.0168 | 0.002096 | 0.0133 | 0.01661 | 0.02136 |
| sigma[1,2] | 0.0027 | 0.003266 | -0.00364 | 0.002736 | 0.008928 |
| sigma[2,1] | 0.0027 | 0.003266 | -0.00364 | 0.002736 | 0.008928 |
| sigma[2,2] | 0.05243 | 0.006214 | 0.0415 | 0.05195 | 0.06583 |

*Figure 3: Posterior kernel densities for the variances and covariance of the random errors*

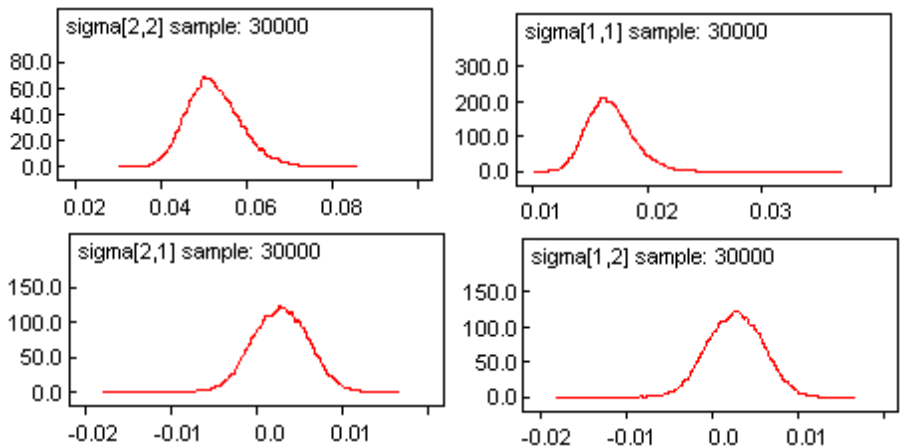

We now get to the main goal of the whole exercise, namely to obtain estimates, in the form of posterior means, for each time period, of the value of the Markov chain on MMDA marketing activity. In Figure 4, the blue line displays the posterior means of states (coded as 0 and 1) for the Markov chain at each time period (between 0 and 1, 0 represents no MMDA marketing activity and 1 some MMDA marketing activity). We also display in Figure 4 (red line) a rescaled variable on actual MMDA marketing expenditures at each time period, in order to evaluate how successful the posterior means are at identifying time periods when marketing activity did take place. It is clear from Table 5 (where we estimated that MMDA marketing activity took place if the posterior mean of the Markov chain state was above 0.5) and Figure 4 that those posterior means do quite a good job of identifying periods with MMDA marketing activity. We will discuss these results further in the next section of the paper.

*Table 5: Competitor Activity Prediction*

| Proportion of classification | |
|---|---|
| Correct identification of the presence of competitor activity (of 58) | 41 (70.69%) |
| Correct identification of the absence of competitor activity (of 98) | 90 (91.84%) |
| Correct identification of the presence or absence of competitor activity (of 156) | 131 (83.97%) |

*Figure 4: Actual and imputed competitor activity*

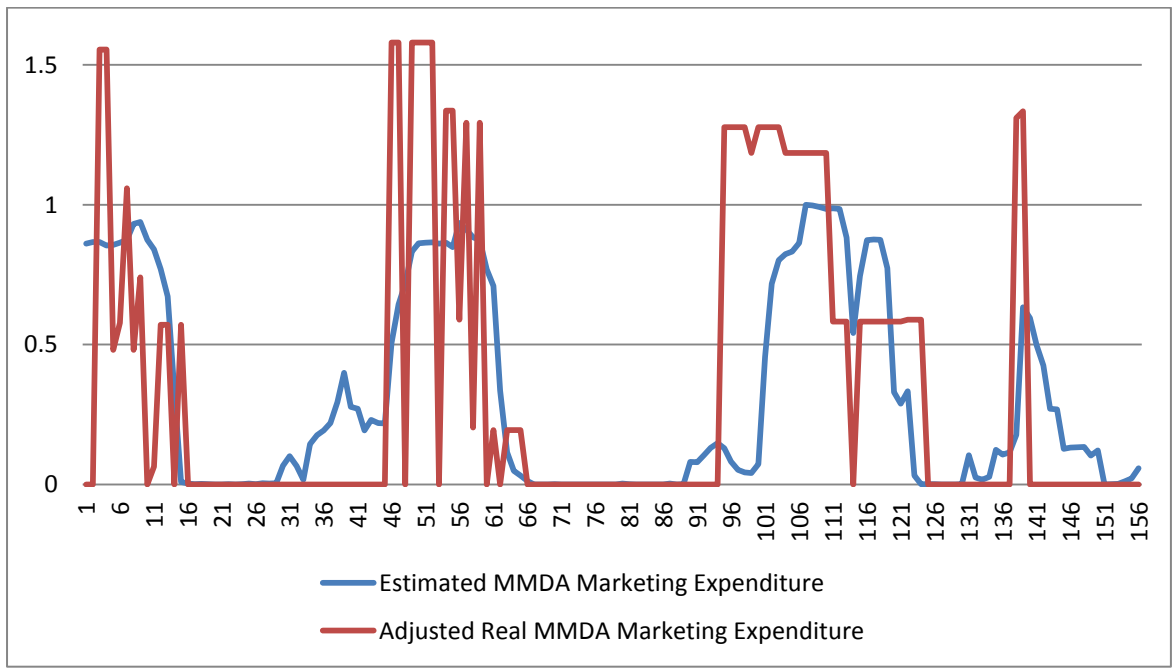

One might wonder whether using a hidden Markov model with more than two states, for example three, gives better results. Interestingly, the model with three states is somewhat less accurate at imputing the missing competitor activity, and is also sensitive to the choice of priors. Moreover, the Winbugs run for estimating this model takes more time. Using three states may over fit the data to some extent.

## Discussion

In deriving the entries of Table 5, we used a cutoff point of 0.5 for the posterior means beyond which we estimated that MMDA marketing activity took place. Of course, as is common to many classification problems, different cutoff points yield different classification results.

Essentially, the intuition behind the results in Table 5 and Figure 4 is that, once models are built for the number of new checking accounts and the number of new MMDA accounts with suitable predictors, unexplained residuals (such as unexplained spikes in competing sales volume) could be attributed to probable competing marketing activity. The latent Markov chain attempts to capture this effect, taking into account that competing marketing activity could very well have an effect on sales volumes of the focal product as well as on those of the competing product.

How would the practitioner use such results? To answer this question, we observe on Figure 4 that the posterior means in general do ferret out the presence of competing marketing activity, but are sometimes a bit fuzzy in their prediction of when exactly the expenditures were put in place. However, they do yield an informative view of competing marketing strategy.

Moreover, without correcting for unknown competitor marketing, the risk is to overestimate the impact of marketing expenditures for the focal brand (represented by the coefficient $\beta_{1c}$ in our equations). In the case presented here, the estimated value of $\beta_{1c}$ is 1.266 (posterior mean) when accounting for competitor marketing activity, and 1.284 without, when marketing expenditures are expressed in million dollars. While this is not a massive difference, it is not negligible, and more importantly, examining the evolution of the impact $\beta_{1c}$ of focal marketing expenditures on the number of new checking accounts is likely to be very informative.

A possible strategy would be to update the model every time period, and to look at the evolving trend followed by the posterior means (blue line) as an indication of competing marketing trends. In particular, such a strategy would yield a picture of the evolution of the impacts $\beta_{1c}$ of focal marketing expenditures on the number of new checking accounts at each model update, in the presence of competing marketing activity.

## Symmetric competitor imputation: can brand B impute promotion activity for Brand A?

An interesting question arises: if each competing agent can predict the unknown marketing activity of the other, and assuming that each competitor can act essentially instantaneously on this knowledge, might it be the case that marketing activities for the two competing parties could cancel each other? This interesting issue would lend itself well to a simulation effort, outside the scope of this paper, to try to see what happens when each side can impute marketing activities for the other and act on it.

We investigated the issue and conducted the same analysis, but switching the roles of checking and MMDA accounts, in other words pretending this time that marketing activity for checking accounts is unknown and that for MMDA accounts is known.

Interestingly, the imputation of the promotional activity for the checking accounts assuming that that of the MMDA accounts is known is not as successful, as can be easily seen in Figure 5. One might say that in this case the imputation of competitor activity is asymmetric. It would be very interesting, from a theoretical point of view, to define and investigate a concept of symmetry of competitor activity imputation; we plan to investigate this issue further in a future paper. Such a discussion would include a review of work on duopolies in economics, going back to ideas by Cournot (*Cournot duopoly equilibrium*).

*Figure 5: Actual and imputed competitor activity, assuming that marketing activity for checking accounts is unknown*

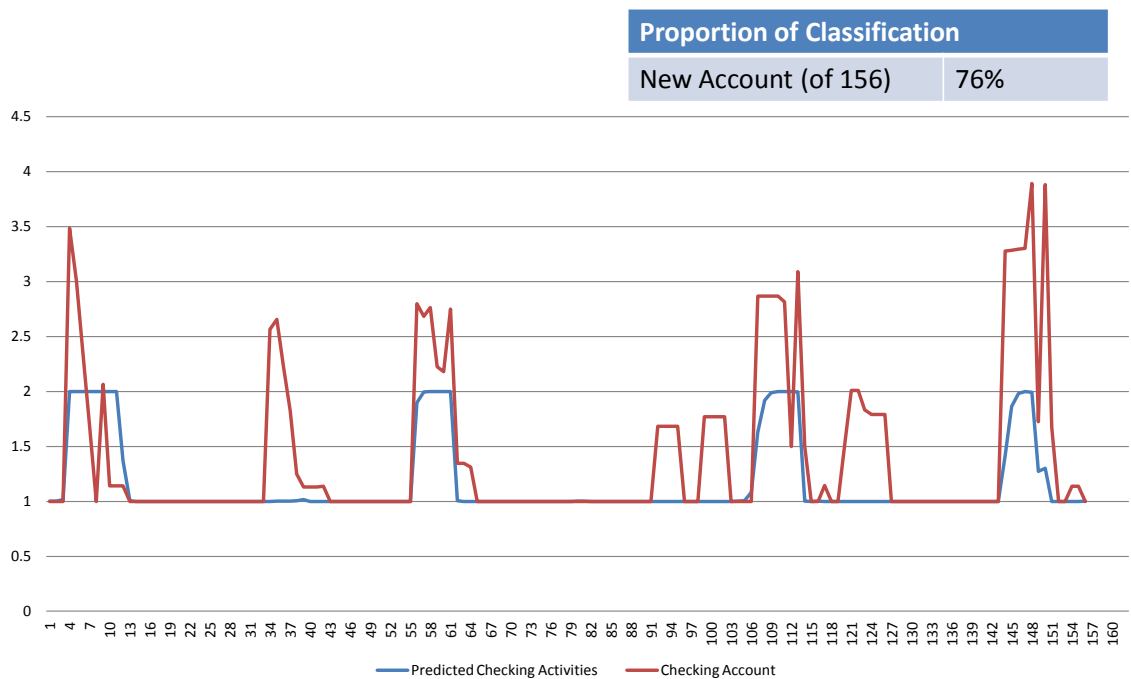

## Conclusion

To conclude, we point out a few directions for further research. It would be useful to test the methodology on a number of other situations where two major products compete. A next step is to build models on the basis of a rolling time window as each new time period is added. This allows one to examine the evolution of the trends in imputed competing marketing activity and to define actionable strategies on that basis. Finally the theoretical concept of symmetry in competitor activity imputation is well worth examining further.

**Appendix: Winbugs code to construct the best model (with partial data listing for confidentiality reasons)**

### HMM Competitor

```
Model
{# latent transition Matrix with 2 latent classes
                x[1] ~ dcat(pin[])
# period 1 latent class
                pin[1:2]   ~ ddirch(mix[]);
# periods >1 latent class
# define the latent variable
for (t in 2:n){ x[t] ~ dcat(P[x[t-1],1:2])}
# define the vector of dependent variables
for (i in 1:n)
{y[i ,1 :2] ~dmnorm(mu[i, 1:2],omega[,])
mu[i,1]<-beta0a+beta1a*z[i]+beta2a*x[i]
mu[i,2]<-beta0b+beta1b*z[i]+beta2b*x[i]}
# define the priors
omega[1:2,1:2]~dwish(r[,],4)
sigma[1:2,1:2]<-inverse(omega[,])
beta0a ~ dnorm(0, 1.0E-6)
beta0b ~ dnorm(0, 1.0E-6)
beta1a ~ dnorm(0, 1.0E-6)
beta1b ~ dnorm(0, 1.0E-6)
beta2a ~ dnorm(0, 1.0E-6)
beta2b ~ dnorm(0, 1.0E-6)

for (j  in 1:2) { for (k in 1:2){  Px[j,k]   ~ dgamma(mix[k],1);
                    P[j,k] <- Px[j,k]/sum(Px[j,])}}

}
```

### Data

```
list(n=156, mix=c(1,1), y = structure(.Data = c(
-0.19976,0.17676,
-0.1761,0.36519,
-0.04936,0.35122,
-0.42536,0.36399,
-0.21376,0.45015,
-0.06772,0.47127,
0.01225,0.46343,
0.22543,0.50913,
0.27634,0.50158,

..... more rows of data ……

.Dim = c(156, 2)),
r = structure(.Data = c(1, 0, 0, 1), .Dim = c(2, 2)),
z=c(
0,
```

```
0,
0,
497011.13,
401075.48,
269753.15,
133372.21,
0,
213177.49,
27889,
27889,
27889,
0,
0,
0,

..... more rows of data ……

0,
0,
27538,
27538,
0
))
```

# Generate Initial values

```
list(beta0a=0,beta0b=0,beta1a=0,beta1b=0,beta2a=0.beta2b=0,pin=c(.5,.5),
x=c(
1,
1,
1,
1,
1,
1,
1,
1,

….. more rows with 1 ….

1,
1,
1,
1,
1),
omega=structure(.Data=c(1,0,0,1),.Dim=c(2,2)),Px=structure(.Data=c(.5,.5,.5,.5),.Dim=c(2,2)) )
```